\documentclass[sigconf,authordraft]{acmart}
\pdfoutput=1

\AtBeginDocument{%
  \providecommand\BibTeX{{%
    \normalfont B\kern-0.5em{\scshape i\kern-0.25em b}\kern-0.8em\TeX}}}

\renewcommand\footnotetextcopyrightpermission[1]{} 

\usepackage{multirow}
\usepackage{graphicx}
\usepackage{hyperref}

\begin{document}

\title{A LLM-based Controllable, Scalable, Human-Involved User Simulator Framework for Conversational Recommender Systems}

\author{Lixi Zhu}
\affiliation{%
  \institution{School of Computer Science and Technology, Beijing Jiaotong University}
  \city{Beijing}
  \country{China}
  \postcode{100044}}
\email{zlxxlz1026@gmail.com}

\author{Xiaowen Huang}
\authornote{corresponding author}
\affiliation{%
  \institution{School of Computer Science and Technology, Beijing Jiaotong University}
  \institution{Beijing Key Lab of Traffic Data Analysis and Mining
, Beijing Jiaotong University}
  \institution{Key Laboratory of Big Data \& Artificial Intelligence in Transportation(Beijing Jiaotong University), Ministry of Education}
  \city{Beijing}
  \country{China}
  \postcode{100044}}
\email{xwhuang@bjtu.edu.cn}

\author{Jitao Sang}
\affiliation{%
  \institution{School of Computer Science and Technology, Beijing Jiaotong University}
  \institution{Beijing Key Lab of Traffic Data Analysis and Mining
, Beijing Jiaotong University}
  \institution{Key Laboratory of Big Data \& Artificial Intelligence in Transportation(Beijing Jiaotong University), Ministry of Education}
  \city{Beijing}
  \country{China}
  \postcode{100044}}
\email{jtsang@bjtu.edu.cn}



\begin{abstract}
Conversational Recommender System (CRS) leverages real-time feedback from users to dynamically model their preferences, thereby enhancing the system's ability to provide personalized recommendations and improving the overall user experience. CRS has demonstrated significant promise, prompting researchers to concentrate their efforts on developing user simulators that are both more realistic and trustworthy. The emergence of Large Language Models (LLMs) has marked the onset of a new epoch in computational capabilities, exhibiting human-level intelligence in various tasks. Research efforts have been made to utilize LLMs for building user simulators to evaluate the performance of CRS. Although these efforts showcase innovation, they are accompanied by certain limitations. In this work, we introduce a \textbf{C}ontrollable, \textbf{S}calable, and \textbf{H}uman-\textbf{I}nvolved (CSHI) simulator framework that manages the behavior of user simulators across various stages via a plugin manager. CSHI customizes the simulation of user behavior and interactions to provide a more lifelike and convincing user interaction experience. Through experiments and case studies in two conversational recommendation scenarios, we show that our framework can adapt to a variety of conversational recommendation settings and effectively simulate users' personalized preferences. Consequently, our simulator is able to generate feedback that closely mirrors that of real users. This facilitates a reliable assessment of existing CRS studies and promotes the creation of high-quality conversational recommendation datasets.
\end{abstract}

\ccsdesc[500]{Information systems~Recommender systems}

\keywords{User Simulator, Conversational Recommender Systems, Large language models}



\maketitle

\section{Introduction}
Recommender systems have significantly enhanced performance and customer satisfaction in various industries. Traditional recommender systems\cite{he2017neural, koren2009matrix, cheng2016wide}, however, primarily analyze users' historical behaviors to model their long-term preferences. This limitation makes it challenging for these systems to address two important questions\cite{gao2021advances}: what is the user's current preference, and which specific aspects of an item does the user favor? Conversational Recommender System (CRS)\cite{sun2018conversational}, typically comprising a dialogue module and a recommendation module, can capture not only the user's long-term preferences but also their real-time preferences. This is achieved through multiple rounds of natural language interactions with the user, enabling the system to make personalized recommendations in real-time. Conversational recommendation has shown considerable promise in daily interactions, where people often seek advice through conversations, such as inquiring about worthwhile movies, dining establishments worth trying, and recently appealing music. Despite the considerable advancements in the field of CRS, The high costs involved in training and evaluating these systems through real human conversations often lead researchers to use user simulators instead. In attribute-based CRS\cite{lei2020estimation, lei2020interactive, deng2021unified, xu2021adapting, zhao2022knowledge}, the user simulator's responses are based on fixed templates, neglecting the flow of the conversation. In contrast, natural language processing (NLP)-based CRS\cite{chen2019towards, zhou2020improving, wang2022barcor, wang2022towards, liang2021learning} takes into account the flow of the conversation, but evaluations are based on fixed conversations, which may neglect the interactivity of conversational recommendations.

Large Language Models (LLMs)\cite{zhao2023survey} have the capability to leverage their extensive world knowledge and commonsense reasoning for text understanding and generation\cite{dai2023uncovering, gao2023chat, wei2022chain, yao2024tree}, in some aspects approaching human-level intelligence\cite{saparov2024testing, xi2023rise, wang2023survey}. The use of LLMs to simulate user behavior has demonstrated encouraging outcomes in areas such as collaboration work\cite{chen2023agentverse, mou2024unveiling} and recommendation systems\cite{wang2023rethinking, wang2023recagent, zhang2023generative}. Recent research have utilized LLMs as user simulators for assessing the performance of CRS\cite{wang2023rethinking, huang2023recommender}, overcoming the template-based limitations of previous research and demonstrating promising results. However, the majority of current studies in constructing user simulators offer session-level guidance to LLMs on generating responses via a single prompt template, which we refer to as ‘single-prompt’. The limitations present in these studies have been analyzed in previous research\cite{zhu2024reliable}, which can be succinctly summarized as data leakage occurs in the user simulator's feedback and controlling the output of the user simulator through a single prompt template proves challenging.

\begin{figure*}[htbp]
    \centering
    \includegraphics[width=0.95\textwidth]{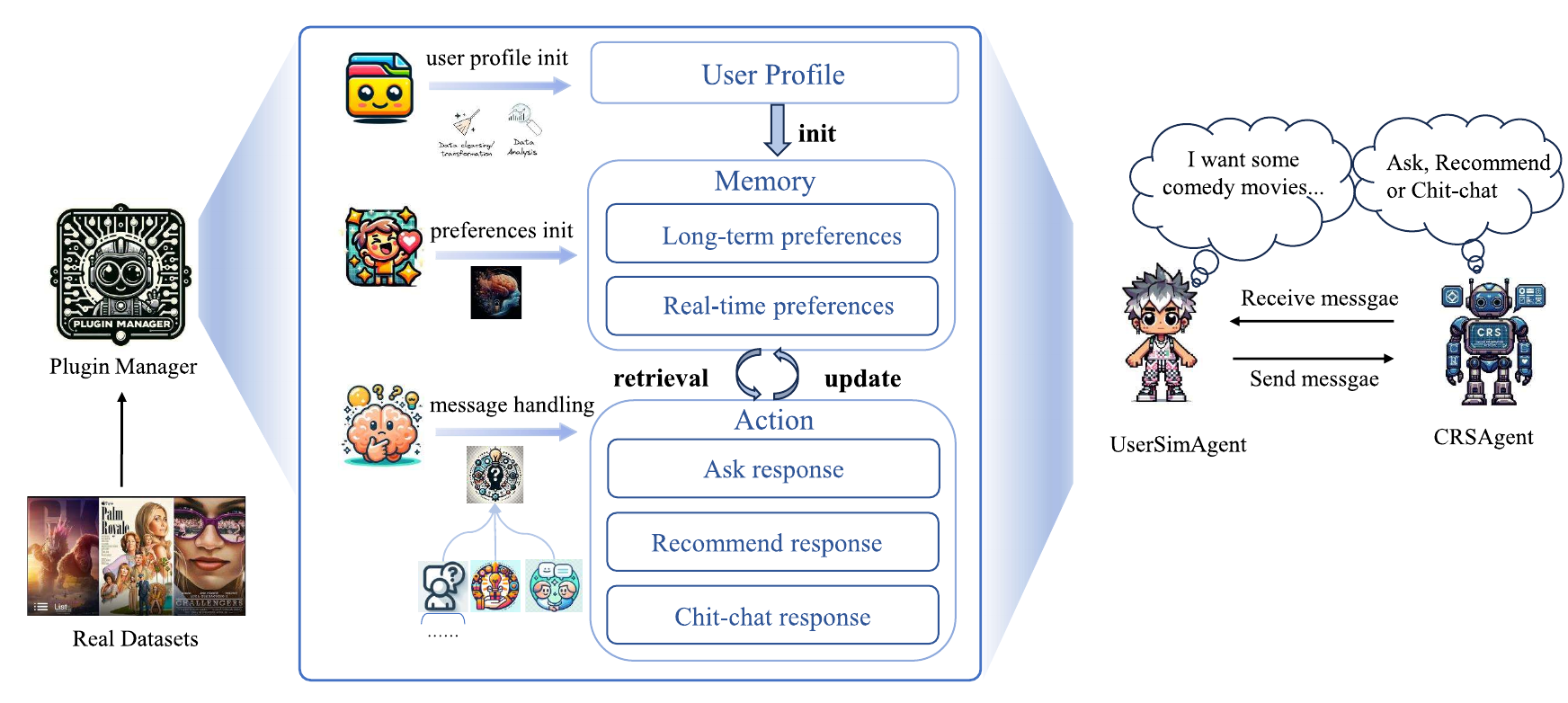}
    \caption{CSHI framework overview.}
    \label{fig-2}
\end{figure*}

In this work, we propose a LLM-based Controllable, Scalable and Human-Involved (CSHI) simulator framework that manages the behavior of user simulators across various stages via a plugin manager. The controllability of CSHI is demonstrated by its precise ability to manage the behavior of the user simulator through a plugin-based design. The scalability of CSHI is evident in its modular design, allowing for the expansion or reduction of plugins and stages to accommodate personalized requirements. Given the complexity and variety of real-world situations, and the distinct operational modes these scenarios demand, the stages of CSHI are designed to be flexible and extendable. The human-involved characteristic allows humans to edit the
profiles of the CSHI-based user simulator agents, influencing their behavior and enabling direct participation in interactions with the CRS. This facilitates the guidance of conversation topics, such as simulating a user’s desire to find a related movie through a memory of a specific movie poster. CSHI customizes the simulation of user behavior and interactions to provide a more lifelike and convincing user interaction experience. 


In summary, our contributions are as follows:
\begin{itemize}
    \item We propose a LLM-based Controllable, Scalable and Human-Involved (CSHI) simulator framework that addresses key issues present in existing user simulators, such as data leakage and the difficulties associated with controlling feedback using a single prompt template. CSHI-based user simulator agents are more realistic and closely mirror real-world scenarios.
    \item Through experiments and case studies in two conversational recommendation scenarios, we show that CSHI-based user simulator agent can adapt to a variety of conversational recommendation settings and effectively simulate users' personalized preferences.
    \item Our simulator is expected to be used to generate high-quality conversational recommendation datasets, thereby advancing research in this field. We have implemented a web-based demo that allows real users to interact with our simulator. It can be accessed at \url{https://github.com/zlxxlz1026/CSHI}.
\end{itemize}



\section{PROPOSED FRAMEWORK}
The overall framework of CSHI is illustrated in Figure \ref{fig-2}. CSHI manages the behavior of user simulator agents across various stages via a plugin manager. Plugins are modular components that can be easily added or removed, offering a plug-and-play functionality that allows for swift customization and adaptation to meet specific needs. This enables the simulator to flexibly handle a variety of complex scenarios and provide a more realistic user interaction experience. Within CSHI, the tasks of each stage are accomplished through the division and collaboration of various plugins. The execution of tasks can either be independently carried out by a single plugin or achieved through the cooperative effort of multiple plugins, among which priorities can be set as needed. In our initial setup, the construction of CSHI-based user simulator agent is divided into three stages: (1) User Profile Init; (2) Preferences Init; (3) Message Handling. In the first stage, User Profile Init, CSHI generates user profiles that are further utilized in the second stage, Preferences Init, to establish the agent's long-term preferences. During the second tage, Preferences Init, CSHI constructs the agent’s preference-related memory, which encompasses both long-term and real-time preferences. In the third stage, Message Handling, the agent retrieves relevant preferences from its memory in response to messages from the CRS and executes corresponding actions. If there are updates to the preferences during this stage, these are recorded back into the memory.

Compared to existing user simulators, CSHI-based user simulator agent introduces three notable improvements: (1) To prevent the problem of data leakage in the feedback from the user simulator, CSHI-based agent adopts a strategy that simulates real user memory. Specifically, before recommending the target item, the agent's memory does not include the specific name of the target item; (2) To generate more realistic user feedback, we draw on the distinction between search and recommendation, to further refine real-time preferences into two categories:  known preferences and unknown preferences. Known preferences refer to those that users can explicitly express, such as wanting to watch a comedy movie from the 90s. Unknown preferences, on the other hand, refer to those that users are not yet aware of or cannot explicitly express. These preferences need to be guided and elicited through information provided by the CRS. For example, a user may initially be unaware of their preference for a particular director's work, but when the CRS introduces relevant movie or director information (such as presenting movies directed by Stephen Chow), the user may then display a clear preference for that director. The discovery and clarification of such preferences occur gradually through the information guidance provided by the recommendation system and the process of user interaction. (3) Sensitive information is anonymized. Sensitive information refers to domain-specific details closely related to the item, such as the duration of a movie or its release date. Users rarely express their preferences in terms such as "I want to find a movie that is 132 minutes long" or "I want to watch a movie that was released on June 1, 2012";

Next, we will provide a phased overview of the plugins contained within each stage of the CSHI-based user simulator agent.


\subsection{User Profile Init}
User profiles play a crucial role in ensuring the consistency of agent behavior. The CSHI offers two methods for generating profiles at this stage: (1) Manual configuration: To enhance agent personalization, CSHI supports manual editing of profiles by users to design agents with distinct personalities. For instance, traits such as "You are very patient and will patiently answer every question" or "You are impatient and prefer to conclude conversations quickly" can be added to construct agents with varied personalities. Although manual editing of profiles offers high flexibility and can endow agents with any characteristics, it may become time-consuming and laborious when the requirements are complex and the agents are diverse. (2) LLMs generation: At this stage, plugins can automate the generation of different information in profiles, for example, extracting a user's long-term taste preferences based on their interaction history, or adding basic information such as the user's interaction history, gender, age, etc. This method can efficiently generate profiles, making it particularly suitable for scenarios that require generic information.
\subsubsection{User Preferences Summary Plugin:}
This plugin, inspired by Agent4Rec\cite{zhang2023generative}, employs a LLM to summarize users' personalized preferences and rating patterns in the movie domain from their interaction history. The approach involves categorizing movies rated 3 and above as "liked" by the user, while those rated below 3 are considered "disliked". This process can be represented as follows:
\begin{equation}T_u=plugin_1(R_u) \end{equation}
Where $R_u$ represents user $u$ ratings for a collection of movies, and the output of the plugin is the personalized preferences $T_u$ of user.

\subsubsection{Basic Information Configuration Plugin:}
This plugin processes raw data to generate basic configuration information for users, such as age, gender, interaction history, etc. This process can be described as follows:
\begin{equation}P_u=plugin_2(RD_u)\end{equation}
Where $RD_u$ represents the raw data of user $u$ in the dataset, and the output of the plugin is the configuration information $P_u$ of the user.

\subsection{Preferences init}
In recommendation systems, user preferences are typically categorized into long-term preferences and real-time preferences. Long-term preferences are modeled based on the user's interaction history, reflecting the user's consistent preferences; whereas real-time preferences denote the user's short-term inclinations, which may align with or differ from their long-term preferences. For example, a user who generally favors science fiction movies might suddenly wish to watch a comedy when feeling down. CRS can capture these real-time feedbacks through interactions with users, facilitating more accurate personalized recommendations. Therefore, at this stage, CSHI generates memories for the user simulator agent that include both long-term and real-time preferences, based on the user profile and target items. Real-time preferences involve information related to target items, while long-term preferences, shaped by the user profile, reflect the user's personalized characteristics.
\subsubsection{Real-Time Preference Generation Plugin:}
This plugin is tasked with generating real-time preferences for the agent during the initialization phase of the user simulator. Its design comprises two main components: the segmentation of real-time preferences and the anonymization of sensitive information. In terms of preference segmentation, the plugin differentiates between "known preferences" and "unknown preferences." Anonymization of sensitive information refers to the handling of domain-specific information related to items, such as changing a movie's specific release date from "June 1, 2012," to "the 2010s" or altering the runtime from "144 minutes" to "about 2 hours." This process can be summarized as follows:
\begin{equation}real\_time\_preference = Plugin_3(Info_i,k_1,k_2)\end{equation}
Where $Info_i$ represents the relevant information generated by the LLMs for target items, such as the movie's director, actors, genre, and language. The hyperparameters $k_1$ and $k_2$ are used to adjust the proportion of known preferences and unknown preferences within the agent, respectively. This plugin generates the user's real-time preferences and integrates them into the agent's memory.
\subsection{Message handling}
This stage is focused on facilitating the interactions between the user simulator and the CRS, with the aim of producing responses that closely resemble human feedback. Upon receiving a message from the CRS, the agent emulates human cognitive processes, first understanding the intent of the CRS, and then executing corresponding actions based on the intent. Within our framework, for common conversational scenarios, the user simulator principally performs three types of actions: ask response, recommend response, and chit-chat response.
\begin{itemize}
    \item Ask response: When the agent is asked about its preferences, it provides suitable answers by retrieving information from its memory.
    \item Recommend response: Faced with item recommendations, the agent gives feedback based on its own preferences.
    \item Chit-chat response: The agent can generate natural and smooth replies based on the current conversation topic and skillfully guide the conversation back to the recommendation context when needed.
\end{itemize}
\subsubsection{Intent Understanding Plugin:}
Upon receiving a message from the CRS, it discerns the system's intent, identifying whether the message is asking about the user's preferences, making a recommendation, or simply engaging in chit-chat. If the message involves an ask, the plugin further identifies the specific content of the ask, such as which aspect of preferences is being asked about. This process can be illustrated as follows:
\begin{equation}intent,rel_{attr}=Plugin_4(message_{last})\end{equation}
Where $message_{last}$ represents the lastest message received from the CRS, this plugin inputs the recognized intent and related queries $rel_{attr}$ into the various response plugins.
\subsubsection{Ask Response Plugin - Personalized Reply:}
When the intent of the CRS is identified as `ask', this plugin is prioritized to initiate, simulating the human memory retrieval process. It first utilizes a LLM to search the user's profile to determine if there exists a set of items relevant to the question asked by the CRS. If relevant items are found, the plugin combines the user's long-term and real-time preferences to generate a personalized reply, thereby concluding the ask response process. Conversely, if no relevant items are retrieved, the process moves to a non-personalized reply plugin. This process can be illustrated as follows:
Where $profile_u$ represents the user's profile, and $flag$ is a Boolean indicator. If $flag$ is true, it signifies that the long-term preferences are unrelated to the current question, resulting in $ask_{response}$ being null, necessitating a transition to the non-personalized reply plugin. Conversely, $ask_{response}$ constitutes the agent's feedback to the CRS's question.
\subsubsection{Ask Response Plugin - Non-Personalized Reply:}
This plugin responds directly to the CRS's question based solely on the agent's real-time preferences. This process can be represented as follows:
\begin{equation}ask_{response}=Plugin_6(intent,real\_time\_preference)\end{equation}
\subsubsection{Recommend Response Plugin:}
When the CRS's intent is identified as `recommend', this plugin initially assesses whether to accept the recommendation. If accepted, it combines previous conversational history to provide positive feedback. If not accepted, the plugin further analyzes the reasons provided by the recommendation system for the recommendation, checking for information related to unknown preferences, If such information is found, it will not only offer negative feedback but also update the agent's real-time preferences based on this information, feeding back this new preference to the CRS. This process can be represented as follows:
\begin{equation}rec_{response},new\_preference=Plugin_7(intent,real\_time\_preference)\end{equation}
Where $rec_{response}$ represents the agent's feedback to the CRS's recommendation. If the real-time preferences are updated, then $new\_preference$ denotes the updated real-time preferences.
\subsubsection{Chit-Chat Response Plugin:}
When the intent of the CRS is identified as `chit-chat‘, this plugin generates smooth and natural replies based on the conversational history. However, considering that the fundamental goal of conversational recommendation is to make recommendations aligned with the user's preferences, this plugin also seamlessly includes its own preferences when generating replies, thereby guiding the direction of the conversation closer to the ultimate recommendation goal. This process can be illustrated as follows:
\begin{equation}chit_{response}=Plugin_8(intent,real\_time\_preference)\end{equation}
Where $chit_{response}$ represents the agent's chit-chat feedback to the CRS.

\section{EXPERIMENTS}
\begin{figure}[htp]
    \includegraphics[width=0.48\textwidth]{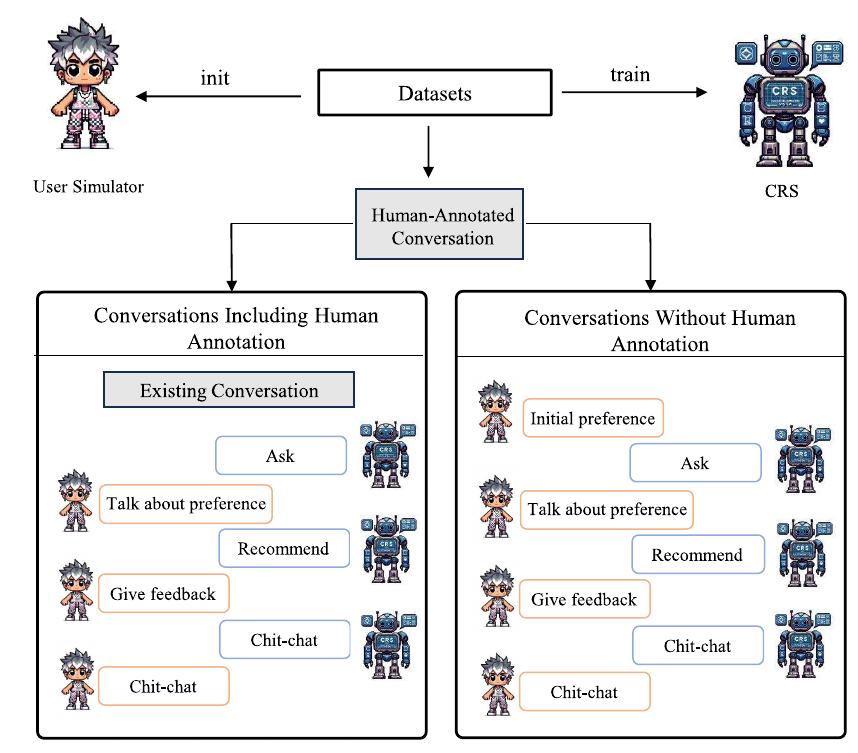}
    \caption{Workflow of the User Simulator.}
    \label{fig-1}
\end{figure}
The workflow of the user simulator is illustrated in Figure \ref{fig-1}. We conducted experiments in two conversational recommendation scenarios, taking into account the characteristics of the dataset.

The first scenario is built upon classic conversational recommendation datasets. These datasets typically contain human-annotated conversations and target item information. The user simulator utilizes the information related to the target item as real-time preferences, interacting with the CRS within predefined conversation contexts until the system successfully recommends an item or reaches the predetermined maximum number of interaction rounds. Here, the simulator's performance might be influenced by these annotated conversations. To address this, we introduced a scenario without annotated conversations, enabling the user simulator and the CRS to embark on entirely new conversations without any pre-established contexts, thus allowing for a more free-form interaction.

The second scenario is based on classic datasets from various fields, such as the MovieLens dataset in the movie recommendation domain. These datasets do not contain human-annotated conversations. The user simulator extracts long-term preferences based on the user's interaction history and treats the information related to the target item as real-time preferences. By integrating long-term and short-term preferences, profiles are generated to guide the user simulator in initiating new conversations with the CRS.
\subsection{Conversational Recommendation Scenarios with Human Annotations}
\subsubsection{Experimental setup}
Our work builds upon iEvaLM\cite{wang2023rethinking}. For detailed information about our experimental settings, please refer to iEvaLM\footnote{https://github.com/RUCAIBox/iEvaLM-CRS/}.
\newline
\textbf{Dataset:} We conduct experiments on two classic datasets in the conversational recommendation domain: ReDial\cite{li2018towards} and OpenDialKG\cite{moon2019opendialkg}. ReDial is a movie conversational recommendation dataset that contains 10,006 conversations. OpenDialKG is a multi-domain conversational recommendation dataset with 13,802 conversations, and only the movie domain was utilized in our experiments.
\newline
\textbf{Baselines:} We conduct comparative experiments using three classical approaches in the field of conversational recommendations, as well as with a well-known large language model named ChatGPT:
\begin{itemize}
    \item KBRD\cite{chen2019towards}: It utilizes an external Knowledge Graph (KG) to enhance the semantics of entities mentioned in the conversational history.
    \item BARCOR\cite{wang2022barcor}: It proposes a unified CRS based on BART\cite{lewis2019bart}, which tackles two tasks using a single model.
    \item UniCRS\cite{wang2022towards}: It proposes a unified CRS model that leverages knowledge-enhanced prompt learning.
    \item ChatGPT \footnote{https://chat.openai.com/}: We employ the publicly available model GPT-3.5-turbo-0613 provided by the OpenAI API.
\end{itemize}
\textbf{Evaluation Metrics:} Following existing work, we adopt Recall@$k$ to evaluate the recommendation task. Similarly, we set $k = 1, 10, 50$, for both the ReDial and OpenDialKG datasets. Additionally, following existing work, we set the maximum number of interaction turns between the user simulator and the CRS at $t = 5$.
\subsubsection{Experimental result}
\begin{figure*}[htbp]
    \centering
    \includegraphics[width=0.95\textwidth]{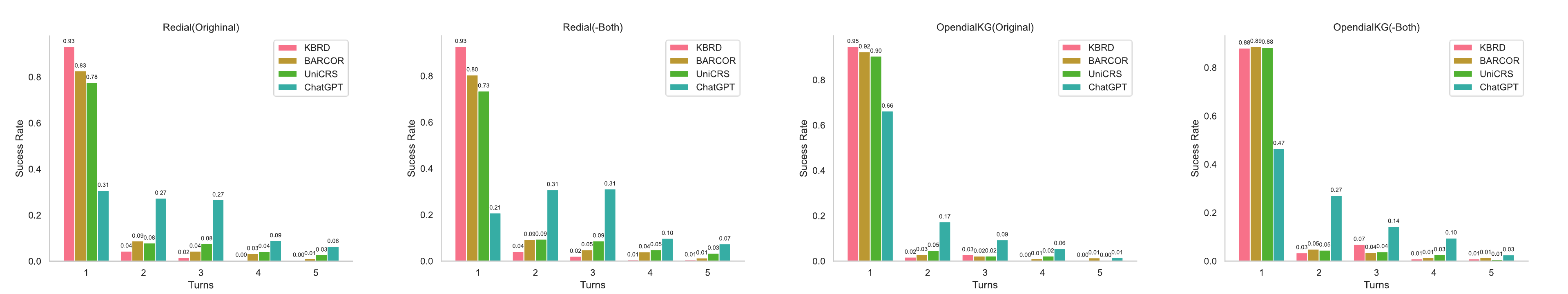}
    \caption{Percentage of successful recommendations by turn when constructing user simulators using CSHI.}
    \label{fig-3}
\end{figure*}

The results of the classic user simulator work, iEvaLM, have already been presented in the charts of reasearch work\cite{zhu2024reliable}. In this experiment, we solely provide the performance of various CRS methods when CSHI replaced iEvaLM in the construction of user simulator. Here, ’-history’ denotes that in our evaluation, conversations influenced by data leakage from conversational history, resulting in successful recommendations, are excluded from consideration. `-response’ denotes that conversations affected by data leakage from the user simulator, which result in successful recommendations, are excluded from consideration. `-both’ denotes that scenarios involving data leakage leading to successful recommendations in both aforementioned contexts are not considered in our evaluation. The evaluation results are presented in Table \ref{table:1} and Figure \ref{fig-3}. We can get the following observations:
\begin{table*}[htb]
\caption{Performance of existing CRS methods and ChatGPT under CSHI in various data leakage scenarios.}
\resizebox{\textwidth}{!}{
\begin{tabular}{cc|ccc|ccc|ccc|ccc}
\hline
\multicolumn{2}{c|}{Model}                                                                   & \multicolumn{3}{c|}{KBRD}                                                                                                                                                                                    & \multicolumn{3}{c|}{BARCOR}                                                                                                                                                                                  & \multicolumn{3}{c|}{UniCRS}                                                                                                                                                                                  & \multicolumn{3}{c}{ChatGPT}                                                                                                                                                                                 \\ \hline
\multicolumn{2}{c|}{datasets}                                                                & Recall@1                                                           & Recall@10                                                          & Recall@50                                                          & Recall@1                                                           & Recall@10                                                          & Recall@50                                                          & Recall@1                                                           & Recall@10                                                          & Recall@50                                                          & Recall@1                                                           & Recall@10                                                          & Recall@50                                                         \\ \hline
\multirow{4}{*}{Redial}     & CSHI                                                            & 0.027                                                              & 0.170                                                              & 0.445                                                              & 0.033                                                              & 0.196                                                              & 0.476                                                              & 0.064                                                              & 0.266                                                              & 0.564                                                              & 0.209                                                              & 0.445                                                              & 0.668                                                             \\
                            & \begin{tabular}[c]{@{}c@{}}CSHI\\ (-history)\end{tabular}       & \begin{tabular}[c]{@{}c@{}}0.009\\ (-66.7\%)\end{tabular}          & \begin{tabular}[c]{@{}c@{}}0.116\\ (-31.8\%)\end{tabular}          & \begin{tabular}[c]{@{}c@{}}0.387\\ (-13.0\%)\end{tabular}          & \begin{tabular}[c]{@{}c@{}}0.034\\ (+0.03\%)\end{tabular}          & \begin{tabular}[c]{@{}c@{}}0.185\\ (-5.6\%)\end{tabular}           & \begin{tabular}[c]{@{}c@{}}0.446\\ (-6.3\%)\end{tabular}           & \begin{tabular}[c]{@{}c@{}}0.024\\ (-62.5\%)\end{tabular}          & \begin{tabular}[c]{@{}c@{}}0.206\\ (-22.6\%)\end{tabular}          & \begin{tabular}[c]{@{}c@{}}0.529\\ (-6.2\%)\end{tabular}           & \begin{tabular}[c]{@{}c@{}}0.183\\ (-12.4\%)\end{tabular}          & \begin{tabular}[c]{@{}c@{}}0.417\\ (-6.3\%)\end{tabular}           & \begin{tabular}[c]{@{}c@{}}0.654\\ (-2.1\%)\end{tabular}          \\
                            & \begin{tabular}[c]{@{}c@{}}CSHI\\ (-response)\end{tabular}      & \begin{tabular}[c]{@{}c@{}}0.027\\ (-0.0\%)\end{tabular}           & \begin{tabular}[c]{@{}c@{}}0.170\\ (-0.0\%)\end{tabular}           & \begin{tabular}[c]{@{}c@{}}0.445\\ (-0.0\%)\end{tabular}           & \begin{tabular}[c]{@{}c@{}}0.033\\ (-0.0\%)\end{tabular}           & \begin{tabular}[c]{@{}c@{}}0.196\\ (-0.0\%)\end{tabular}           & \begin{tabular}[c]{@{}c@{}}0.476\\ (-0.0\%)\end{tabular}           & \begin{tabular}[c]{@{}c@{}}0.064\\ (-0.0\%)\end{tabular}           & \begin{tabular}[c]{@{}c@{}}0.264\\ (-0.8\%)\end{tabular}           & \begin{tabular}[c]{@{}c@{}}0.564\\ (-0.0\%)\end{tabular}           & \begin{tabular}[c]{@{}c@{}}0.209\\ (-0.0\%)\end{tabular}           & \begin{tabular}[c]{@{}c@{}}0.445\\ (-0.0\%)\end{tabular}           & \begin{tabular}[c]{@{}c@{}}0.668\\ (-0.0\%)\end{tabular}          \\
                            & \textbf{\begin{tabular}[c]{@{}c@{}}CSHI\\ (-both)\end{tabular}} & \textbf{\begin{tabular}[c]{@{}c@{}}0.009\\ (-66.7\%)\end{tabular}} & \textbf{\begin{tabular}[c]{@{}c@{}}0.116\\ (-31.8\%)\end{tabular}} & \textbf{\begin{tabular}[c]{@{}c@{}}0.387\\ (-13.0\%)\end{tabular}} & \textbf{\begin{tabular}[c]{@{}c@{}}0.034\\ (+0.03\%)\end{tabular}} & \textbf{\begin{tabular}[c]{@{}c@{}}0.0185\\ (-5.6\%)\end{tabular}} & \textbf{\begin{tabular}[c]{@{}c@{}}0.446\\ (-6.3\%)\end{tabular}}  & \textbf{\begin{tabular}[c]{@{}c@{}}0.024\\ (-62.5\%)\end{tabular}} & \textbf{\begin{tabular}[c]{@{}c@{}}0.204\\ (-23.3\%)\end{tabular}} & \textbf{\begin{tabular}[c]{@{}c@{}}0.529\\ (-6.2\%)\end{tabular}}  & \textbf{\begin{tabular}[c]{@{}c@{}}0.183\\ (-12.4\%)\end{tabular}} & \textbf{\begin{tabular}[c]{@{}c@{}}0.417\\ (-6.3\%)\end{tabular}}  & \textbf{\begin{tabular}[c]{@{}c@{}}0.654\\ (-2.1\%)\end{tabular}} \\ \hline
\multirow{4}{*}{OpendialKG} & CSHI                                                            & 0.243                                                              & 0.432                                                              & 0.558                                                              & 0.271                                                              & 0.412                                                              & 0.532                                                              & 0.256                                                              & 0.456                                                              & 0.609                                                              & 0.403                                                              & 0.690                                                              & 0.900                                                             \\
                            & \begin{tabular}[c]{@{}c@{}}CSHI\\ (-history)\end{tabular}       & \begin{tabular}[c]{@{}c@{}}0.079\\ (-67.5\%)\end{tabular}          & \begin{tabular}[c]{@{}c@{}}0.213\\ (-50.7\%)\end{tabular}          & \begin{tabular}[c]{@{}c@{}}0.353\\ (-36.7\%)\end{tabular}          & \begin{tabular}[c]{@{}c@{}}0.190\\ (-29.9\%)\end{tabular}          & \begin{tabular}[c]{@{}c@{}}0.299\\ (-27.4\%)\end{tabular}          & \begin{tabular}[c]{@{}c@{}}0.383\\ (-28.0\%)\end{tabular}          & \begin{tabular}[c]{@{}c@{}}0.153\\ (-40.2\%)\end{tabular}          & \begin{tabular}[c]{@{}c@{}}0.295\\ (-35.3\%)\end{tabular}          & \begin{tabular}[c]{@{}c@{}}0.456\\ (-25.1\%)\end{tabular}          & \begin{tabular}[c]{@{}c@{}}0.202\\ (-49.9\%)\end{tabular}          & \begin{tabular}[c]{@{}c@{}}0.543\\ (-21.3\%)\end{tabular}          & \begin{tabular}[c]{@{}c@{}}0.877\\ (-2.6\%)\end{tabular}          \\
                            & \begin{tabular}[c]{@{}c@{}}CSHI\\ (-response)\end{tabular}      & \begin{tabular}[c]{@{}c@{}}0.243\\ (-0.0\%)\end{tabular}           & \begin{tabular}[c]{@{}c@{}}0.432\\ (-0.0\%)\end{tabular}           & \begin{tabular}[c]{@{}c@{}}0.558\\ (-0.0\%)\end{tabular}           & \begin{tabular}[c]{@{}c@{}}0.272\\ (+0.4\%)\end{tabular}           & \begin{tabular}[c]{@{}c@{}}0.412\\ (-0.0\%)\end{tabular}           & \begin{tabular}[c]{@{}c@{}}0.531\\ (-0.2\%)\end{tabular}           & \begin{tabular}[c]{@{}c@{}}0.257\\ (+0.4\%)\end{tabular}           & \begin{tabular}[c]{@{}c@{}}0.457\\ (+0.2\%)\end{tabular}           & \begin{tabular}[c]{@{}c@{}}0.608\\ (-0.2\%)\end{tabular}           & \begin{tabular}[c]{@{}c@{}}0.403\\ (-0.0\%)\end{tabular}           & \begin{tabular}[c]{@{}c@{}}0.691\\ (+0.1\%)\end{tabular}           & \begin{tabular}[c]{@{}c@{}}0.900\\ (-0.0\%)\end{tabular}          \\
                            & \textbf{\begin{tabular}[c]{@{}c@{}}CSHI\\ (-both)\end{tabular}} & \textbf{\begin{tabular}[c]{@{}c@{}}0.079\\ (-67.5\%)\end{tabular}} & \textbf{\begin{tabular}[c]{@{}c@{}}0.213\\ (-50.7\%)\end{tabular}} & \textbf{\begin{tabular}[c]{@{}c@{}}0.353\\ (-36.7\%)\end{tabular}} & \textbf{\begin{tabular}[c]{@{}c@{}}0.191\\ (-29.5\%)\end{tabular}} & \textbf{\begin{tabular}[c]{@{}c@{}}0.298\\ (-27.7\%)\end{tabular}} & \textbf{\begin{tabular}[c]{@{}c@{}}0.380\\ (-28.6\%)\end{tabular}} & \textbf{\begin{tabular}[c]{@{}c@{}}0.153\\ (-40.2\%)\end{tabular}} & \textbf{\begin{tabular}[c]{@{}c@{}}0.295\\ (-35.3\%)\end{tabular}} & \textbf{\begin{tabular}[c]{@{}c@{}}0.455\\ (-25.3\%)\end{tabular}} & \textbf{\begin{tabular}[c]{@{}c@{}}0.203\\ (-49.6\%)\end{tabular}} & \textbf{\begin{tabular}[c]{@{}c@{}}0.544\\ (-21.2\%)\end{tabular}} & \textbf{\begin{tabular}[c]{@{}c@{}}0.877\\ (-2.6\%)\end{tabular}} \\ \hline
\end{tabular}}
\label{table:1}
\end{table*}

\begin{itemize}
    \item Data leakage from conversational history is an inherent problem of the dataset. However, we can look forward to the generation of more high-quality conversational recommendation data in the future through the interactions between user simulators and CRS.
    \item  As shown in Table \ref{table:1}, CSHI-based agent significantly mitigates the issue of data leakage caused by user simulators. For instance, as illustrated in analysis work\cite{zhu2024reliable}, when employing iEvaLM as the user simulator, various CRS methods experienced a decline in their recall@50 metric by 12.0\%, 6.8\%, 10.3\%, and 17.9\% on the ReDial dataset. In contrast, when utilizing CSHI to construct user simulator under the same experimental setup, there was no observable decrease in performance.
    \item As illustrated in Figure \ref{fig-3}, in the `-both’ scenario, CSHI-based agent demonstrates superior performance across multiple rounds of interactions (from the 2nd to the 5th round). This superior performance is attributable to the agent’s ability to express its preferences in chit-chat conversational scenarios.
\end{itemize}
\subsection{Conversational Recommendation Scenarios without Human Annotations}
\subsubsection{Experimental setup}
\
\newline
\textbf{Dataset:} We conducted experiments using the classic dataset in the movie recommendation domain, MovieLens. For each user,  we selected their latest five interactions as the test set, while the remaining interaction history were utilized to generate user profiles.
\newline
\textbf{Baselines:} In our experiments, we observed an interesting phenomenon: in conversational recommendation scenarios with human annotations, the task of recommendation is often separate from conversation generation. Therefore, at times, even if the CRS generates target items that match user preferences within the conversation, the system might incorrectly deem the recommendation unsuccessful if the recommendation model does not successfully recommend the item. This is clearly unreasonable. Consequently, in conversational recommendation scenarios without human annotations, we adopted a generative CRS method based on LLMs\cite{li2023large}.

The CRS agent comprises the Strategy Module, Memory Module, and Action Module: (1) The Strategy Module determines the actions to be taken based on the conversation history and historical decision-making behaviors stored in the Memory Module, as well as known user preferences. (2) The Memory Module stores the entire history of the conversation, historical decision-making behaviors, and information on user preferences for retrieval by the Strategy and Action Modules. (3) Actions comprise ask, recommend and chit-chat.
\newline
\textbf{Evaluation Metrics:} Given the challenge for generative CRS to recommend strictly based on item relevance, we utilize the recommendation accuracy at round t (success rate@t, SR@t) to measure the proportion of successful recommendations made by the CRS during the t-th interaction round in this scenario. We use the average number of turns (average turn, AT) to assess the average number of interactions between the user simulator and the CRS within a single conversation. Considering the context length limitations of generative CRS, we set the maximum number of recommended items generated by the CRS agent to 10. In the scenario without human annotations, we set the maximum number of interaction rounds between the user simulator and the conversational recommendation system to $T=10$.
\subsubsection{Experimental result}
\begin{figure}[htp]
    \includegraphics[width=0.45\textwidth]{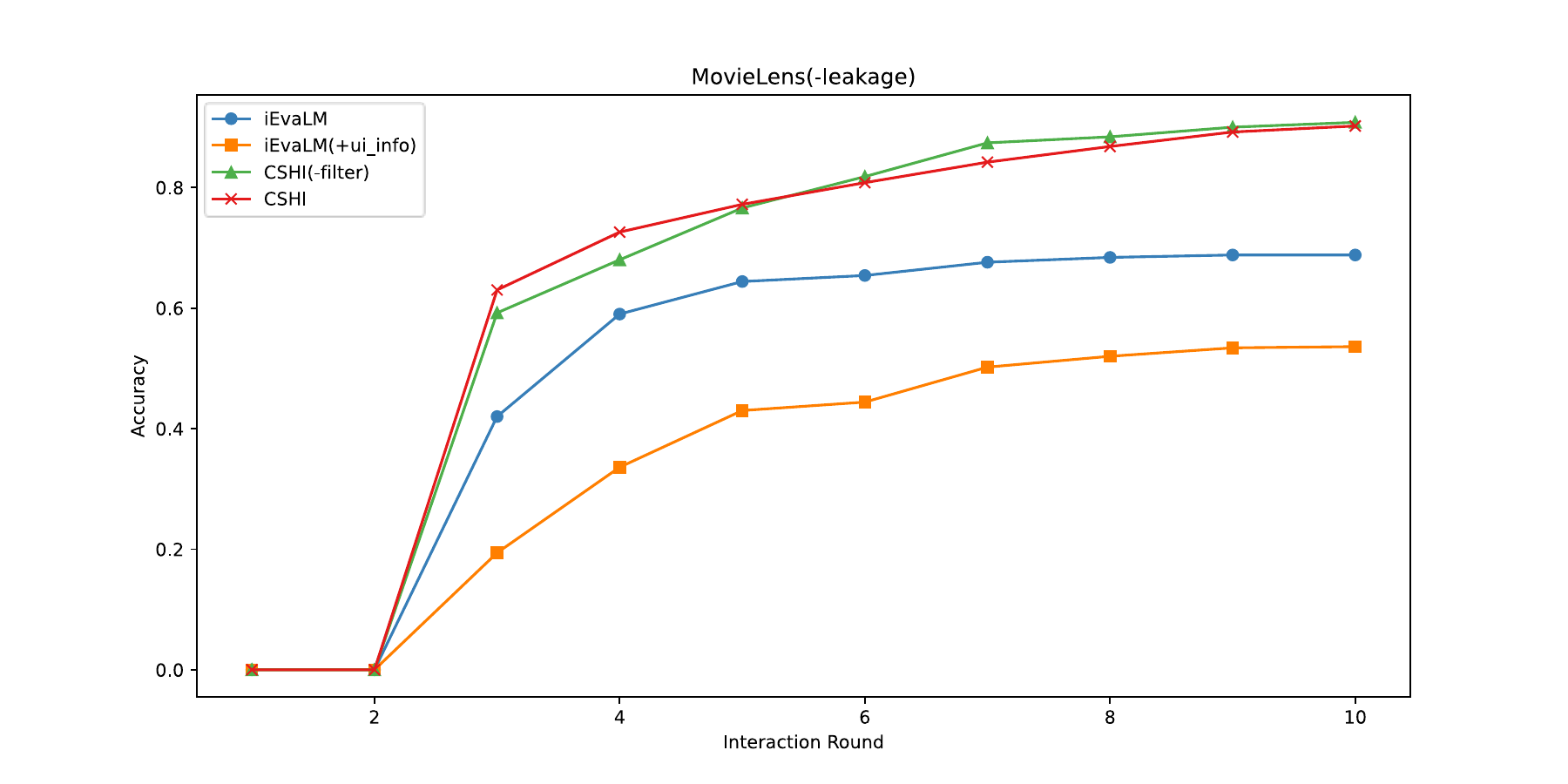}
    \caption{Recommendation accuracy across different rounds.}
    \label{fig-4}
\end{figure}

We conducted comparative and ablation experiments to evaluate the performance of CSHI, iEvaLM, and their variants. Here, "iEvaLM (+ui\_info)" refers to the addition of target item information and user interaction history to the original prompt template. "CSHI (-filter)" indicates that the CSHI-based agent does not consider the anonymization of sensitive information in its processing. "-leakage" is used to denote the performance of the CRS without considering the impact of data leakage on successful conversations. The experimental results are presented in Table \ref{table:2}. Figure \ref{fig-4} displays the recommendation accuracy of the CRS under different user simulator configurations and across various rounds, disregarding successful conversations affected by data leakage. Through the analysis of Figure \ref{fig-4} and Table \ref{table:2}, we can draw the following conclusions:
\begin{table}[]
\caption{Performance of CRS under CSHI in conversational recommendation scenarios without Human Annotations.}
\begin{tabular}{c|cccc}
\hline
                                                                         & SR@3                                                              & SR@5                                                              & SR@10                                                             & AT             \\ \hline
iEvaLM                                                                   & 0.434                                                             & 0.680                                                             & 0.742                                                             & 5.384          \\ \hline
\begin{tabular}[c]{@{}c@{}}iEvaLM\\ (-leakage)\end{tabular}              & \begin{tabular}[c]{@{}c@{}}0.42\\ (-3.2\%)\end{tabular}           & \begin{tabular}[c]{@{}c@{}}0.644\\ (-5.3\%)\end{tabular}          & \begin{tabular}[c]{@{}c@{}}0.688\\ (-7.3\%)\end{tabular}          & 5.104          \\ \hline
iEvaLM+ui\_info                                                          & 0.2                                                               & 0.484                                                             & 0.658                                                             & 6.598          \\ \hline
\begin{tabular}[c]{@{}c@{}}iEvaLM+ui\_info\\ (-leakage)\end{tabular}     & \begin{tabular}[c]{@{}c@{}}0.194\\ (-3.0\%)\end{tabular}          & \begin{tabular}[c]{@{}c@{}}0.43\\ (-11.2\%)\end{tabular}          & \begin{tabular}[c]{@{}c@{}}0.536\\ (-18.5\%)\end{tabular}         & 5.82           \\ \hline
CSHI                                                                      & 0.634                                                             & 0.776                                                             & 0.906                                                             & 4.434          \\ \hline
\textbf{\begin{tabular}[c]{@{}c@{}}CSHI\\ (-leakage)\end{tabular}}        & \textbf{\begin{tabular}[c]{@{}c@{}}0.63\\ (-0.6\%)\end{tabular}}  & \textbf{\begin{tabular}[c]{@{}c@{}}0.772\\ (-0.5\%)\end{tabular}} & \textbf{\begin{tabular}[c]{@{}c@{}}0.902\\ (-0.4\%)\end{tabular}} & \textbf{4.422} \\ \hline
CSHI-filter                                                               & 0.598                                                             & 0.772                                                             & 0.918                                                             & 4.43           \\ \hline
\textbf{\begin{tabular}[c]{@{}c@{}}CSHI-filter\\ (-leakage)\end{tabular}} & \textbf{\begin{tabular}[c]{@{}c@{}}0.592\\ (-1.0\%)\end{tabular}} & \textbf{\begin{tabular}[c]{@{}c@{}}0.766\\ (-0.8\%)\end{tabular}} & \textbf{\begin{tabular}[c]{@{}c@{}}0.908\\ (-1.1\%)\end{tabular}} & \textbf{4.386} \\ \hline
\end{tabular}
\label{table:2}
\end{table}
\begin{itemize}
    \item Compared to iEvaLM, when utilizing CSHI to construct user simulator, the CRS exhibits superior performance in terms of both recommendation accuracy and average number of interaction turns. This advantage stems from the plugin-based working mechanism of CSHI, which renders the output of the user simulator agent more controllable. Meanwhile, single-prompt user simulators sometimes fail to generate feedback that meets expectations.
    \item Both CSHI and iEvaLM(+ui\_info) incorporate the user's interaction history into their memory, yet iEvaLM(+ui\_info) underperforms compared to iEvaLM. Upon case study, we hypothesize that this is due to the inclusion of the user's interaction history and target item information in the prompt template, which leads to an increase in template length and makes the output of the LLMs more challenging to control. A clear indication of this is that despite the prompt template explicitly stating not to mention the name of the target item, the issue of data leakage with iEvaLM(+ui\_info) is more severe than with iEvaLM. This suggests that the addition of information may have a negative impact on the stability and controllability of the user simulator's output.
    \item Compared to CSHI, CSHI(-filter) demonstrates improved performance with an increase in the number of interaction turns. This phenomenon occurs because the CRS tends to request sensitive information such as release dates and durations to later stages. Consequently, if sensitive information is not anonymized, the CRS can leverage these detailed insights provided by the user simulator to make more accurate recommendations. However, this approach does not align with realistic conversational scenarios.
    \item As evident from Figure \ref{fig-4}, the recommendation accuracy in the first two rounds is zero across different interactions between user simulators and the CRS. This indicates that the CRS tends to prefer detailed asks about user preferences in the initial stages to make rapid recommendations subsequently. In later interaction rounds, the recommendation accuracy of iEvaLM gradually stabilizes, suggesting that the feedback provided by iEvaLM offers limited assistance to the CRS as interactions progress. In contrast, CSHI-based agent exhibits a trend of consistent and steady improvement in recommendation accuracy with an increase in interaction rounds, indirectly highlighting CSHI's significant advantage in terms of controllability.
\end{itemize}

From the experimental results in two conversational recommendation scenarios, we observed that CSHI, compared to single-prompt user simulators, can generate more controllable feedback, thereby enhancing the performance of the CRS.
\section{Case Study}
Some advantages of CSHI are not readily apparent through quantitative experiments alone. In this section, we will showcase these benefits and discuss them in detail from the perspectives of controllability, scalability, and human involvement. The following cases are based on real interactions.
\subsection{Controllability}
The controllability of CSHI is primarily manifested in two aspects: the initialization of user preferences and the generation of feedback.
\subsubsection{Feedback Generation}
\begin{figure}[htp]
    \includegraphics[width=0.5\textwidth]{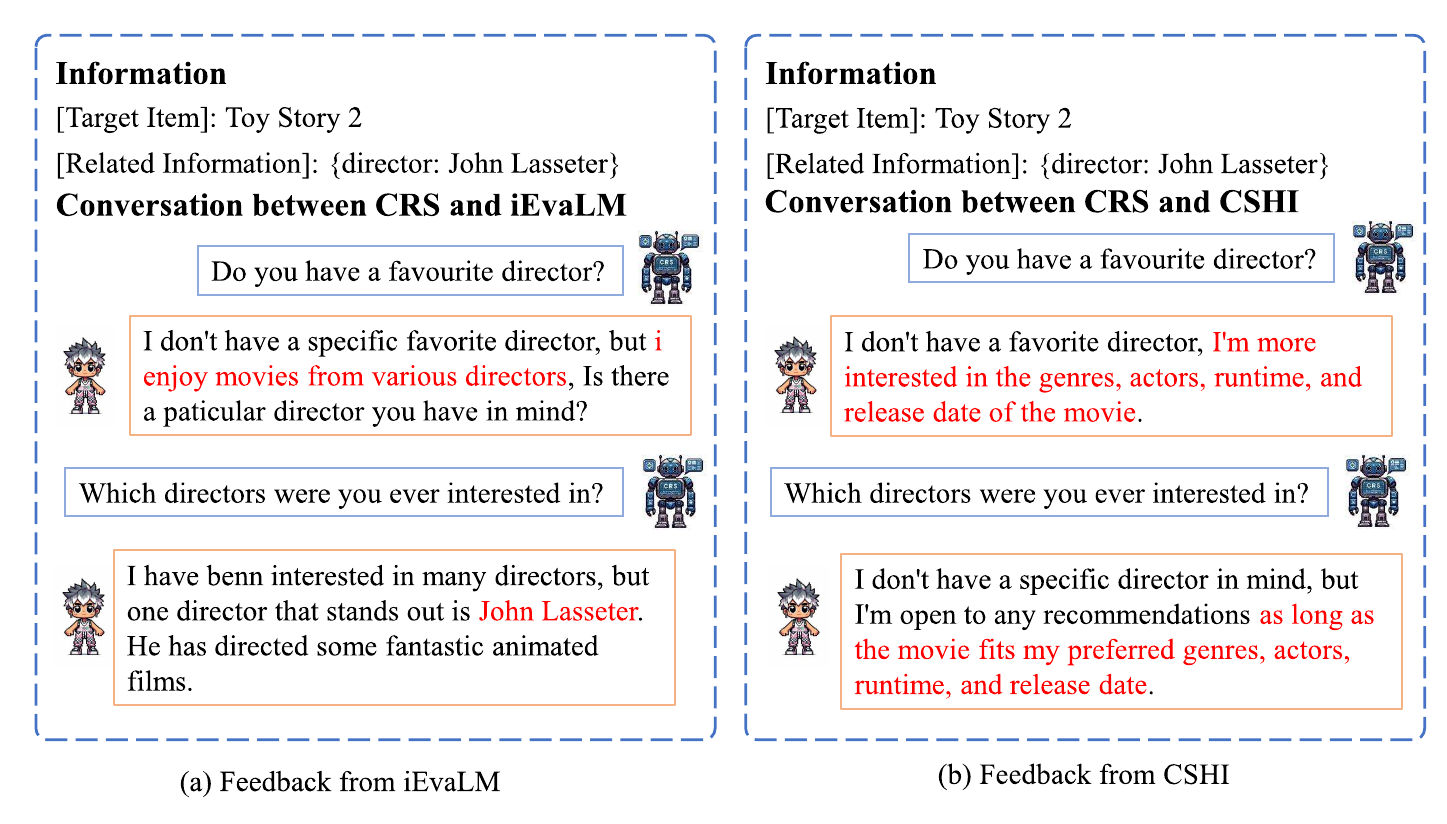}
    \caption{Feedback from Different Simulators.}
    \label{fig-5}
\end{figure}

From Figure \ref{fig-5}, it can be observed that when iEvaLM is asked about preferences related to directors, based on its real-time preferences, it might provide information relevant to the question or might not, with no control over which specific information should be recognized as known preferences and supplied to the CRS. In contrast, CSHI is acutely aware of its real-time preferences. Therefore, when faced with asks about director preferences, CSHI will explicitly indicate a lack of preferences related to directors, emphasizing its focus on aspects such as movie genres and actors instead. 
\subsubsection{Data Leakage}
\begin{figure}[htp]
    \includegraphics[width=0.5\textwidth]{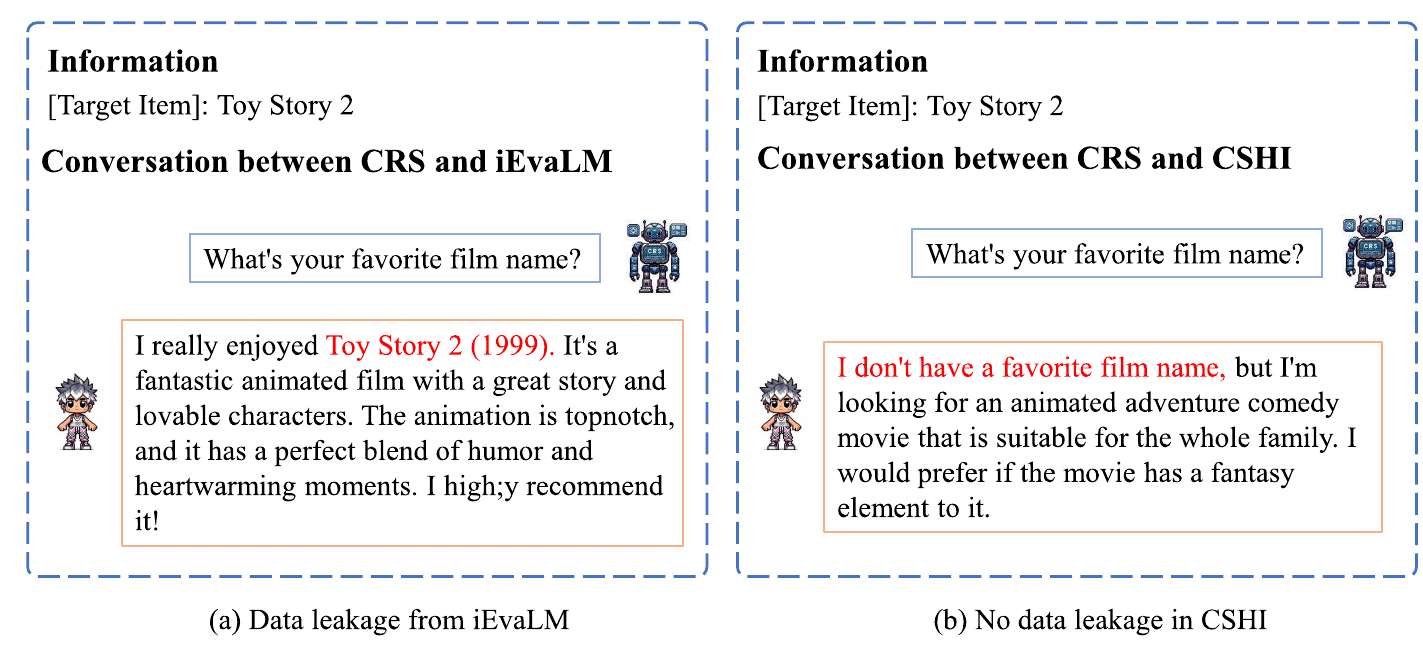}
    \caption{Data leakage from user simulators.}
    \label{fig-6}
\end{figure}

From Figure \ref{fig-6}, it can be observed that even though iEvaLM's prompt template explicitly instructs to avoid disclosing the name of the target items, inadvertent data leakage can still occur due to the inclusion of the target item's name in the template itself. In contrast, as CSHI's memory does not contain any names of target items, it effectively prevents the leakage of target item names.
\subsubsection{Unknown Preferences}
\begin{figure}[htp]
    \includegraphics[width=0.5\textwidth]{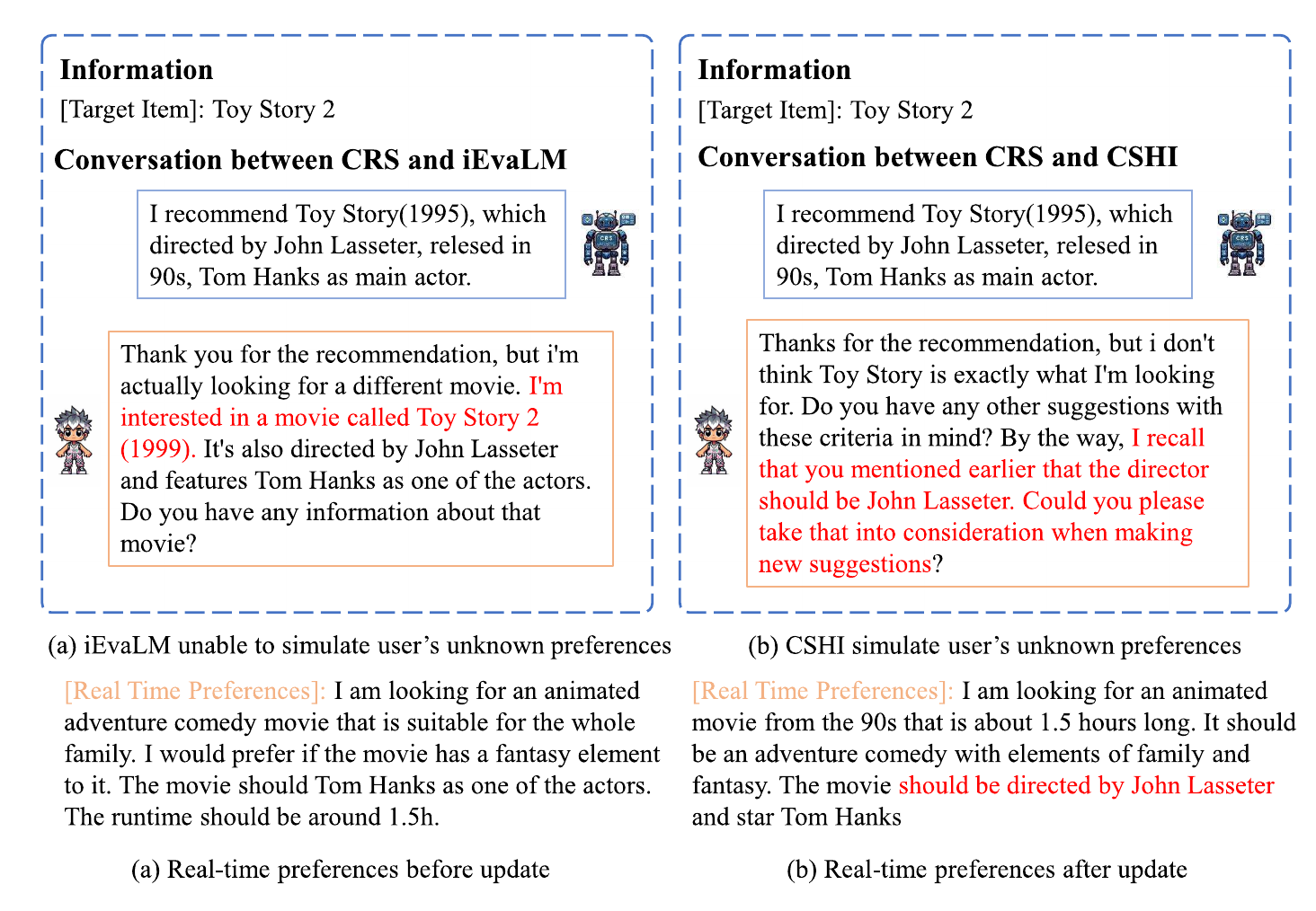}
    \caption{Unknown preferences from user Simulators.}
    \label{fig-7}
\end{figure}

As demonstrated in the Figure \ref{fig-7}, single-prompt user simulators struggle to effectively simulate users' latent preferences. The prevalent approach is to provide information about the target item while rejecting the recommendation list. In contrast, when the CRS provides information related to unknown preferences, CSHI responds positively and updates its real-time preferences.
\subsection{Scalability}
The scalability of CSHI is demonstrated by its ability to meet specific needs through the design of dedicated plugins, such as information anonymization and generating personalized responses by incorporating user interaction history.
\subsubsection{Anonymization of Sensitive Information}
\begin{figure}[htp]
    \includegraphics[width=0.5\textwidth]{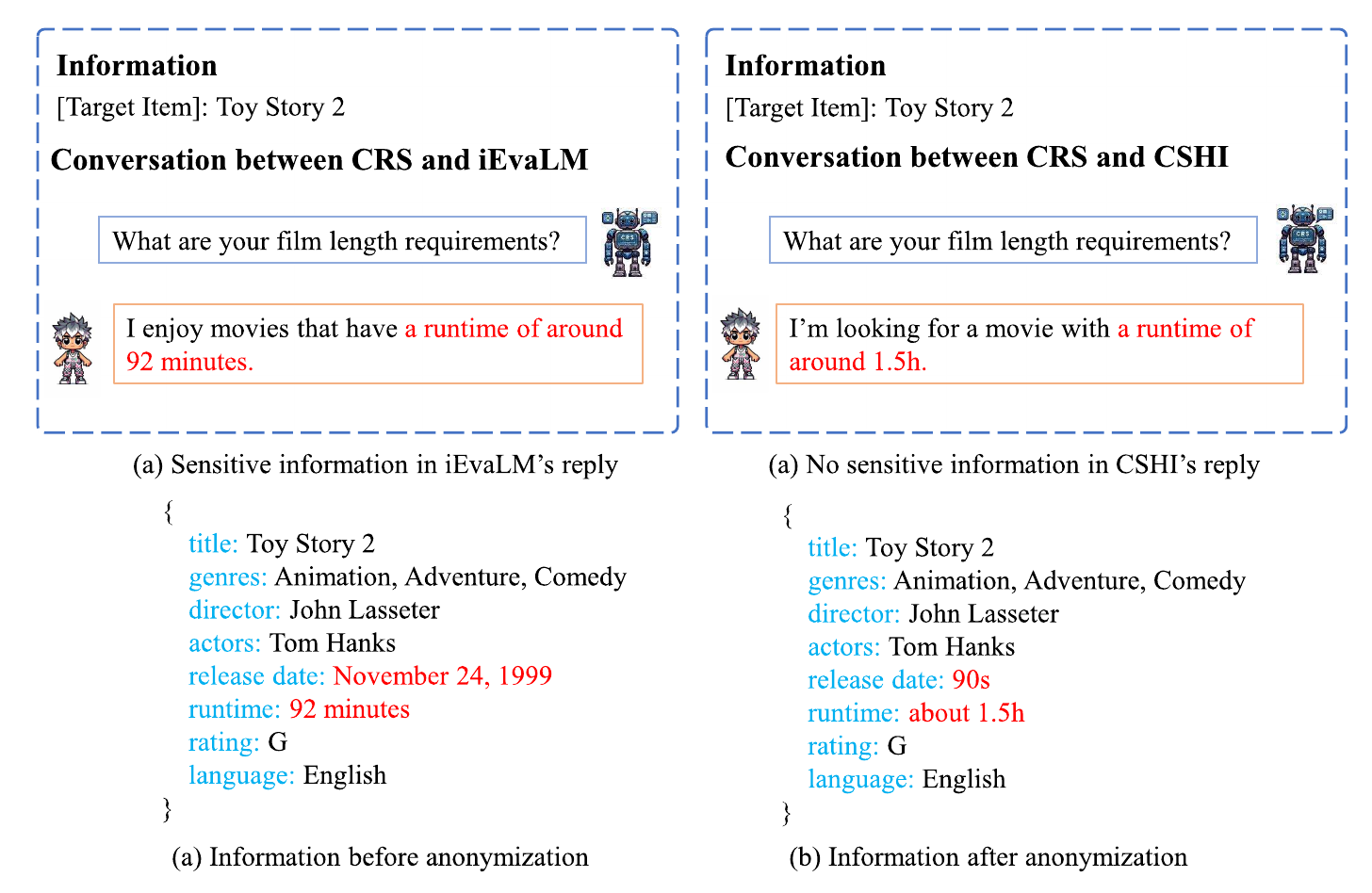}
    \caption{Comparison before and after information anonymization.}
    \label{fig-9}
\end{figure}
From Figure \ref{fig-9}, it is evident that without the anonymization of sensitive information, the user simulator would output domain-specific information related to items, such as a movie's release date. However, in real-life scenarios, it is uncommon for users to express their preferences in such specific terms as "I want to watch a movie released on June 1, 2012."
\subsubsection{Personalized Reply}
\begin{figure}[htp]
    \includegraphics[width=0.5\textwidth]{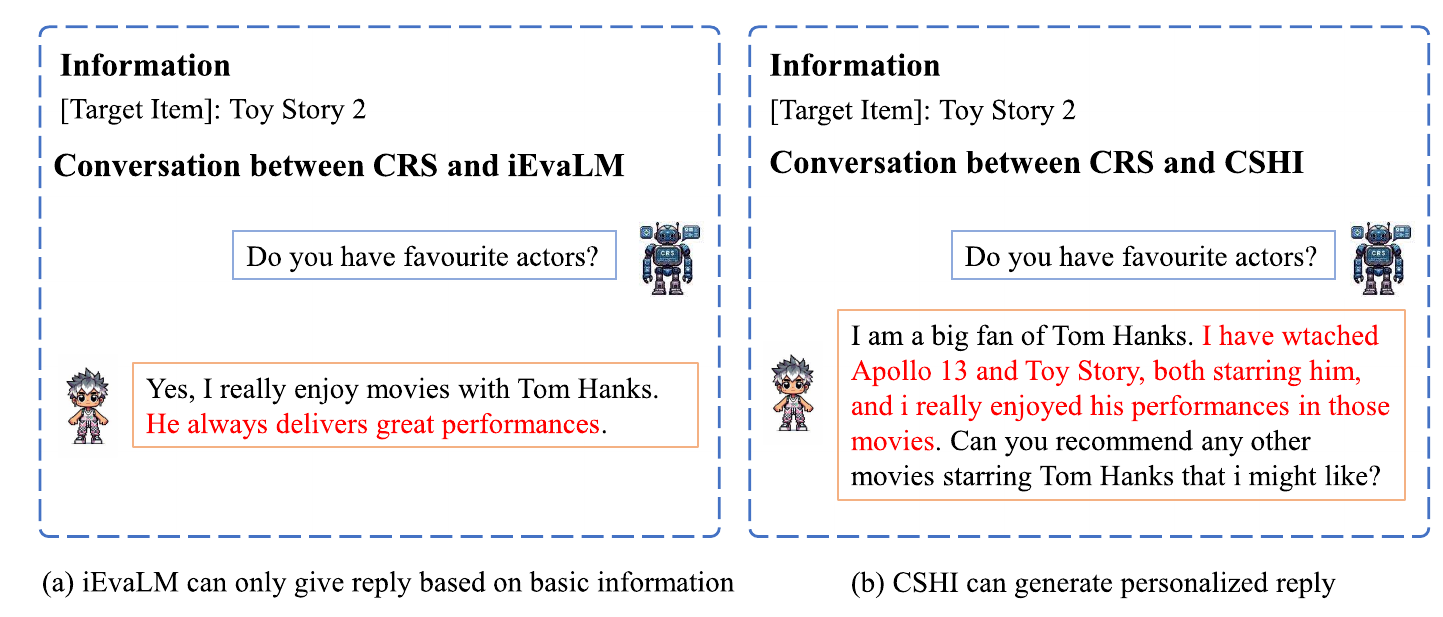}
    \caption{User simulator's personalized reply.}
    \label{fig-11}
\end{figure}

In real-world scenarios, users often favor certain items because they have interacted with similar items previously. Hence, we implemented a plugin that can generate personalized responses by incorporating users' interaction histories. From Figure \ref{fig-11}, it is observable that since different users have distinct interaction histories, the feedback generated, even for the same theme (such as directors), is personalized. iEvaLM struggles to achieve this level of personalization.
\subsection{Human Involvement}
In CSHI, humans can edit the configuration profile of the user simulator agent, thereby influencing the agent's behavior, and can directly engage in interactions with the CRS to guide the conversation topic. This is exemplified by a user wanting to find a related movie through a memory associated with a movie poster.
\begin{figure}[htp]
    \includegraphics[width=0.3\textwidth]{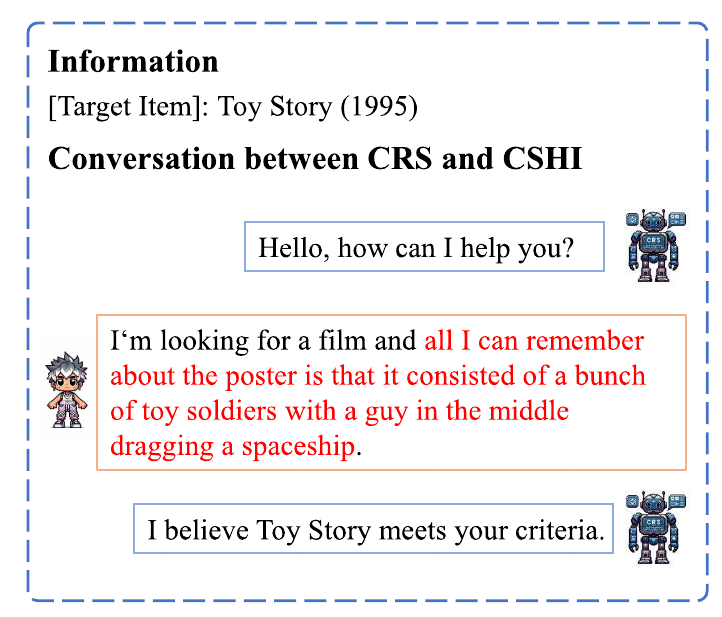}
    \caption{Human involvement in the conversation.}
    \label{fig-12}
\end{figure}

As Figure \ref{fig-12} illustrated, human participants can act as the user's memory by incorporating relevant information about a poster into the user simulator's profile, or they can directly participate in the conversation by providing partial poster information during the interaction, guiding the CRS towards successful recommendations.
\section{CONCLUSION}
In this paper, wo introduced a LLM-based Controllable, Scalable, Human-Involved user simulator framework that manages the behavior of user simulators across various stages via a plugin manager. CSHI customizes the simulation of user behavior and interactions to provide a more lifelike and convincing user interaction experience. Through experiments and case studies in two conversational recommendation scenarios, we show that our framework can adapt to a variety of conversational recommendation settings and effectively simulate users' personalized preferences. CSHI is expected to be used to generate high-quality conversational recommendation datasets, thereby advancing research in this field.

Currently, we have only attempted to manually input visual features, such as movie posters, into agents. Given the scalability of the CSHI, it can conveniently utilize information from various modalities, including visual, to construct user simulation agents. This approach allows for a closer approximation to the behaviors of real users.

\bibliographystyle{unsrt}
\bibliography{sample-base}

\end{document}